\documentclass[prb,showpacs, aps,twocolumn]{revtex4-1}
\usepackage{graphicx}
\usepackage{amsmath}
\usepackage{amssymb}

\begin{document}

\title{Magneto-transport in impurity-doped few-layer graphene spin valve}
\author {Kai-He Ding$^1$, Zhen-Gang Zhu$^2$, Zhen-Hua Zhang$^1$, and Jamal Berakdar$^2$}
\affiliation{$^1$Department of Physics and Electronic Science,
 Changsha University of Science and Technology,
 Changsha,410076, China\\
 $^2$Institut f\"{u}r Physik Martin-Luther-Universit\"{a}t
Halle-Wittenberg Nanotechnikum-Weinberg, Heinrich-Damerow-Strasse
4 D - 06120 Halle (Saale), Germany}


\begin{abstract}
Using Keldysh nonequilibrium Green's function method we study  the
spin-dependent transport through impurity-doped few layer graphene
sandwiched between two magnetic leads with
  an arbitrary mutual orientations of the magnetizations. We find for parallel electrodes
 magnetizations that the
differential conductance  possesses two resonant peaks  as  the applied bias increases.
These peaks are traced
back to a buildup   of a magnetic moment on the impurity  due to the
electrodes spin polarization. For a large mutual angle of the electrodes magnetization directions,
 the two resonant peaks approach each others and merge into
 a single peak for  antiparallel orientation of the  electrodes magnetizations.
We point out that the tunneling magnetoresistance (TMR) may change
sign  for relatively  small changes in the values of the
polarization parameters. Furthermore, we inspect the
behaviour of the
differential conductance and TMR  upon  varying the temperature.
\end{abstract}

 \maketitle
\section{Introduction}
In the past few years much effort was devoted to the
investigations, fabrications and utilizations of  graphene samples
\cite{novoselov2,zhang,geim,castrocond}
 culminating in a series of  fascinating findings such
as the anomalous quantized Hall effect, the absence of weak
localization and the existence of minimal conductivity\cite{geim}. These phenomena
underline the
 remarkable potential of graphene in future advances in nanoscience. Fueled by this development
novel graphene-based technological applications and
 devices are currently envisaged. In this respect, a special attention
 has been paid to  the spin dependent transport in  graphene and graphene-ferromagnet
 heterostructures, as prototypical spintronic device
 \cite{hill2006ieee,tombros2007nature,cho2007apl,ohishi2007apl,
 wang2008prb,ding2009prb,stephan,ding_epl,chen,dingjpcm2008}.
For example, Hill \emph{et al. }\cite{hill2006ieee} fabricated graphene spin valves
 and observed a 10\% change in the resistance as the
electrodes magnetizations switch orientation from a parallel to an antiparallel configuration. Recent
experiments on the spin injection in a single layer graphene show a
rather long spin-flip relaxation length $\approx 1 \mu m$ at room
temperature \cite{tombros2007nature}. The spin injection into a
graphene thin film has been successfully demonstrated by using
nonlocal magnetoresistance
measurements \cite{tombros2007nature,cho2007apl,ohishi2007apl}.
Wang \emph{et al.} \cite{wang2008prb} measured the magnetoresistance of
mesoscopic graphite spin-valve devices and observed a cusplike
feature of the magnetoresistance versus the applied bias. Ding \emph{et
al.} \cite{ding2009prb} studied theoretically the spin-dependent
transport through the graphene spin valve device, and pointed out
that a pronounced cusplike feature at zero bias is due to the
result of a subtle combined effect of graphene and the
conventional spin-valve properties.

Recently, it has been demonstrated \cite{meyer} that
adatoms can be precisely positioned on graphene. Numerous works evidence that the
adatoms may create new many-body states
 in graphene and lead to extraordinary properties such as
magnetism
\cite{Uchoaprl2008,{netossc2008},{dingjpcm2009},{uchoaprl2009}}
and Kondo effect
\cite{hentschelprb2007,{doraprb2007},{cornprl2009},{zhuangepl2009},{zhuarxiv2009},{wehlarxiv2009}},
which are different from the case of impurities in an ordinary
metal. Therefore, the modification of the properties of
graphene by the impurity atoms may also influence
 the magneto-transport  in graphene nanojunctions, an issue which is addressed
here.
 In this work, we study theoretically the spin
dependent transport through  few layers of  graphene in the
presence of impurity atoms. The method is  based on the standard
Keldysh nonequilibrium Green's function approach,
 as described in \cite{haug,rammer}. We find that if the
electrodes spin polarization vectors are parallel,   the applied bias dependence
of the differential conductance exhibits two resonant peaks
signaling the formation of the impurity magnetic moment due to the
electrodes magnetization.
With increasing the mutual angle  $\theta$ of the electrodes spin polarization vectors
the two peaks develop gradually into a single peak at  $\theta=\pi$.
Thus the electrodes magnetization orientations  may be used  to
switch on and off the magnetism \textit{of} the impurity atoms. We
also investigate the dependence of the tunnel-magnetoresistance on
$\theta$, on the temperature, and on the spin polarization degrees
of the electrodes in details.
\section{Single layer garaphene}
\begin{figure}[h]
\includegraphics[width=0.9\columnwidth ]{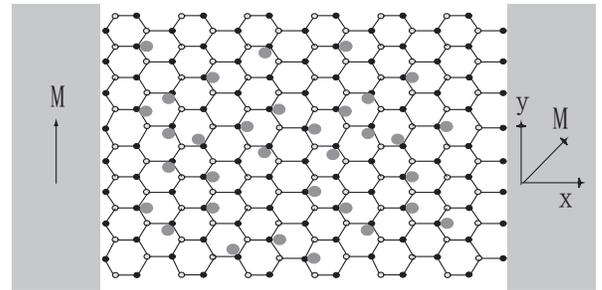}
\caption{A schematic illustration of the system considered in this
work. A nanotunnel junction  made of  one or few layers of
graphene with impurity atoms is connected to two magnetic leads.
The magnetic moments of the leads are aligned by a relative angle
$\theta $, and the coupling matrix between $\alpha (\alpha =L,R)$\
electrode and graphene is $T_{k\alpha }$. }\label{fig1}
\end{figure}
We consider a monolayer graphene sandwiched between two
ferromagnetic electrodes. Some impurity atoms are absorbed on the
top of carbon atoms in the graphene sheet, as shown in
Fig.\ref{fig1}. The moment $\mathbf{M}_L$ of the left electrode is
assumed to align along the $y$ direction, while the moment
$\mathbf{M}_R$ of the right electrode deviates from the $y$
direction by a relative angle $\theta$. A bias voltage $V$ is
applied between the left and the right electrodes. The electric
current flows in the $x$ direction. The Hamiltonian of this system
reads
\begin{equation}
H=H_L+H_R+H_G+H_T+H_i+H_f.
\end{equation}
Here, $H_L(H_R)$ describes the left(right) electrode:
\begin{equation}
H_L=\sum\limits_{\mathbf{k},\sigma} \varepsilon_{\mathbf{k}L\sigma}
c_{\mathbf{k}L\sigma}^\dag c_{\mathbf{k}L\sigma}
\end{equation}
\begin{equation}
H_{R}=\sum\limits_{\mathbf{k},\sigma }[\varepsilon _{R}(\mathbf{k})-\sigma \mathbf{M}%
_{R}\cos \theta ]c_{\mathbf{k}R\sigma }^{\dag
}c_{\mathbf{k}R\sigma }-\mathbf{M}_{R}\sin \theta c_{\mathbf{k}R\sigma
}^{\dag }c_{\mathbf{k}R\overline{\sigma }}  \label{03}
\end{equation}
where $ \varepsilon_{\mathbf{k}\alpha\sigma}$ is the single electron
energy, $c_{\mathbf{k}\alpha\sigma}^\dag(c_{\mathbf{k}\alpha\sigma})$ is the usual
creation (annihilation) operator for an electron with the momentum
$\mathbf{k}$ and the spin $\sigma$ in the $\alpha=L,R$ electrode.

The tight-binding Hamiltonian of the electrons in graphene is
given by
\begin{equation}
H_G=-t\sum\limits_{\langle i,j\rangle,\sigma} (a_{i,\sigma}^\dag
b_{j,\sigma}+\text{H.c.}), \label{hg}
\end{equation}
where $a_{i,\sigma}^\dag (a_{i,\sigma})$ creates (annihilates) an
electron with the spin $\sigma$ on the position $\mathbf{R}_i$ of
the sublattice $A$, $b_{i,\sigma}^\dag (b_{i,\sigma})$
creates(annihilates) an electron with the spin $\sigma$ on the
position $\mathbf{R}_i$ of the sublattice $B$, $t$ is the nearest
neighbor$(\langle i,j\rangle)$ hopping energy. In the momentum
space Eq.(\ref{hg}) can be rewritten as
\begin{equation}
H_G=\sum\limits_{\mathbf{q},\sigma}
[\phi(\mathbf{q})a_{\mathbf{q}\sigma}^\dag
b_{\mathbf{q}\sigma}+\phi(\mathbf{q})^* b_{\mathbf{q}\sigma}^\dag
a_{\mathbf{q}\sigma} ]\label{hg12}
\end{equation}
where $\phi(\mathbf{q})=-t\sum\limits_{i=1}^3
e^{i\mathbf{q}\cdot\mathbf{\delta_i}}$ with
$\delta_1=\frac{a}{2}(1,\sqrt{3},0),
\delta_2=\frac{a}{2}(1,-\sqrt{3},0),\delta_3=a(1,0,0)$(here $a$ is
the lattice spacing). The Hamiltonian (\ref{hg12}) can be easily
diagonalized and one can show $E_{\pm}(\mathbf{q})=\pm
t|\phi(\mathbf{q})|$, which can be linearized
 around the $\mathbf{K}$
points of the Brillouin zone and have a dispersion given by
\begin{equation}
E_{\pm}(\mathbf{q})=\pm v_F|\mathbf{q}|,
\end{equation}
where $v_F=3ta/2$ is the Fermi velocity of an electron.
The coupling between the electrodes and
graphene is modeled by
\begin{equation}
H_T=\frac{1}{\sqrt{N}}\sum\limits_{\mathbf{kq}\alpha\sigma}
[T_{\mathbf{k}\alpha\mathbf{q}}c_{\mathbf{k}\alpha\sigma}^\dag
a_{\mathbf{q}\sigma}+ \text{H.c.}],
\end{equation}
where $T_{\mathbf{k}\alpha\mathbf{q}}$ is the coupling matrix
between the $\alpha$ electrode and the graphene; N is the number
of sites on the sublattice $A$. Assuming  some impurity atoms to
be placed on the top of the sublattice $A$, for the simplicity, we
neglect the correlated interaction between the impurities, and
express the hybridization with the localized impurity states as
\begin{equation}
H_i=\frac{1}{\sqrt{N}}\sum\limits_{\mathbf{q}l\sigma}V_l(f_{l\sigma}^\dag
a_{\mathbf{q}\sigma}+a_{\mathbf{q}\sigma}^\dag f_{l\sigma}),
\end{equation}
where $f_{l\sigma}^\dag (f_{l\sigma})$ creates(annihilates) an
electron with spin $\sigma$ at the impurity at the site $l$, and
$V_l$ is the random and strength distributions of the
hybridization satisfying $\langle
V_lV_{l'}\rangle_{dis}=V_0^2\delta_{ll'}$, where
$\langle\cdots\rangle_{dis}$ denotes the impurity
average\cite{sinitsynprl2006,nunnerprb2007}.

$H_f$ describes the impurities:
\begin{equation}
H_f=\sum\limits_{l\sigma} \varepsilon_0 f^\dag_{l\sigma}
f_{l\sigma}+U n_{l\uparrow} n_{l\downarrow},\label{hf1}
\end{equation}
where $n_{l\sigma}=f_{l\sigma}^\dag f_{l\sigma}$ is the occupation
number operator, $\varepsilon_0$ is the single electron energy at
the impurity, and the Coulomb interaction is included by a finite
$U$ Anderson term. For simplicity, we adopt a mean field
approximation to the electronic correlations at the impurity, $U
n_{l\uparrow} n_{l\downarrow} \simeq U\sum\limits_\sigma \langle
n_{l\overline{\sigma}}\rangle f_{l\sigma}^\dag
f_{l\sigma}-U\langle n_{l\uparrow}\rangle \langle
n_{l\downarrow}\rangle$, meaning that the present theory is
reliable for temperatures above the Kondo temperatures.  The
impurity Hamiltonian is rewritten as $H_f=\sum\limits_\sigma
\varepsilon_{l\sigma} f^\dag_{l\sigma} f_{l\sigma}$ with
$\varepsilon_{l\sigma}=\varepsilon_0+U\langle
n_{l\overline{\sigma}}\rangle$.

The electric current can be calculated from the time evolution of
the occupation number operator of the left electrode.
\begin{equation}
I=e\langle\dot{ \mathcal{N}_L}\rangle=\frac{ie}{\hbar}\langle
[H,\mathcal{N}_L]\rangle, \label{jl}
\end{equation}
where
$\mathcal{N}_L=\sum\limits_{\mathbf{k}\sigma}c_{\mathbf{k}L\sigma}^\dag
c_{\mathbf{k}L\sigma}$. Using the nonequilibrium Green's function
method, Eq.(\ref{jl}) can be further expressed as
\begin{equation}
\begin{array}{cll}
I &=&-\frac{ie}{\hbar }\int\frac{d\varepsilon}{2\pi} Tr \{
[\mathcal{G}_{a}^{r}(\varepsilon)-\mathcal{G}_{a}^{a}(\varepsilon)]
f_L(\varepsilon)
+\mathcal{G}_{a}^{<}(\varepsilon)\}\Gamma_{L}(\varepsilon),
\end{array} \label{jl3}
\end{equation}
where $Tr$ is the trace in the spin space, $f_\alpha(\varepsilon)$
is Fermi distribution function,
$\mathcal{G}_{a}^{r}(\varepsilon)=\sum\limits_{\mathbf{qq'}}\langle
G_{\mathbf{q}a,\mathbf{q'}a}^{r}(\varepsilon)\rangle_{dis}$ and
$\mathcal{G}_{a}^{<}(\varepsilon)=\sum\limits_{\mathbf{qq'}}\langle
G_{\mathbf{q}a,\mathbf{q'}a}^{<}(\varepsilon)\rangle_{dis}$ are  $2\times 2$
matrices representing  the retarded green's function and the
lesser Green's function respectively. In the calculation of
Eq.(\ref{jl3}), we assume that the dominant contributions to
tunneling stem from the electrons near Fermi level, and hence
assume the linewidth function to be independent of $\mathbf{q}$.
Thus, we have
\begin{equation}
\Gamma_{\alpha }=\left(
\begin{array}{cc}
\Gamma_{\alpha }^\uparrow & 0 \\
0 & \Gamma_{\alpha }^\downarrow%
\end{array}%
\right)
\end{equation}
with $\Gamma_\alpha^\sigma =2\pi\sum\limits_\mathbf{k}
T_{\mathbf{k}\alpha\mathbf{q}}^*T_{\mathbf{k}\alpha\mathbf{q'}}
\delta(\varepsilon-\varepsilon_{\mathbf{k}\alpha\sigma})$.

The lesser Green function $\mathcal{G}_{a}^{<}(\varepsilon )$\ can
be calculated by the Keldysh equation
$\mathcal{G}_{a}^{<}(\varepsilon )=\mathcal{G}_{a}^{r}(\varepsilon
)\Sigma ^{<}(\varepsilon )\mathcal{G}_{a}^{a}(\varepsilon )$. To
obtain $\Sigma ^{<}(\varepsilon )$, we invoke Ng's ansatz
\cite{ngprl1996}: $\Sigma ^{<}(\varepsilon )=\Sigma
_{0}^{<}(\varepsilon )B$, where $\Sigma _{0}^{<}(\varepsilon
)=i[\Gamma _{L}f_{L}(\varepsilon )+R\Gamma _{R}R^{\dag
}f_{R}(\varepsilon )]$ is the  lesser self-energy of the clean
graphene system with
$$R=\left(
\begin{array}{cc}
\cos \frac{\theta }{2} & -\sin \frac{\theta }{2} \\
\sin \frac{\theta }{2} & \cos \frac{\theta }{2}%
\end{array}%
\right) .
$$
$B$ is determined by the condition $B=(\Sigma _{0}^{r}(\varepsilon
)-\Sigma _{0}^{a}(\varepsilon
))^{-1}(\mathcal{G}_{a}^{a-1}(\varepsilon
)-\mathcal{G}_{a}^{r-1}(\varepsilon ))$ with $\Sigma
_{0}^{r}(\varepsilon )(\Sigma _{0}^{a}(\varepsilon ))$ denoting
the retarded(advanced) self-energy of the clean graphene system.
Under these considerations, we finally obtain
\begin{equation}
I=\frac{e}{\hbar }\int \frac{d\varepsilon }{2\pi }Tr[\Gamma
_{L}\mathcal{G}_{a}^{r}(\varepsilon )\overline{\Sigma
}^{<}\mathcal{G}_{a}^{a}(\varepsilon )][f_{R}(\varepsilon
)-f_{L}(\varepsilon )],  \label{jL5}
\end{equation}%
where $\overline{\Sigma }^{<}=R\Gamma _{R}R^{\dag }B$. The
remaining task is to calculate the retarded Green's function
$\mathcal{G}_{a}^{r}(\varepsilon )$.

By the equation of motion and then impurity average, we can derive
\begin{equation}
\mathcal{G}_{a}^{r}(\varepsilon)
=\overline{g}_{a}^{r}(\varepsilon)[1-\overline{g}_{a}^{r}(\varepsilon)\Sigma^r(\varepsilon)
]^{-1}, \label{faa2}
\end{equation}
where
$\overline{g}_{a}^{r,a}(\varepsilon)=\frac{1}{N}\sum\limits_\mathbf{q}
g_{\mathbf{q}a,\mathbf{q}a}^{r,a}(\varepsilon)$ and $
\Sigma^r=n_iV_0^2g_c -\frac{i}{2}[\Gamma_{L}(\varepsilon) +R
\Gamma_{R}(\varepsilon)R^\dag] $ with $n_i$ the impurity density,
$g_{\mathbf{q}a,\mathbf{q}a}^{r,a}(\varepsilon)=\frac{\varepsilon}{(\varepsilon\pm
i\eta)^2-|\phi(\mathbf{q})|^2}$, and
\begin{equation}
g_c=\left(
\begin{array}{cc}
\frac{1}{\varepsilon-\varepsilon_\uparrow}&0\\
0&\frac{1}{\varepsilon-\varepsilon_\downarrow}
\end{array}\right).\label{selgc}
\end{equation}
 Introducing a cutoff $k_c$ leads to
\begin{equation}
\overline{g}_{a}^{r}(\varepsilon)=-F_0(\varepsilon)-i\pi\rho_0(\varepsilon),\label{gari}
\end{equation}
\begin{equation}
F_0(\varepsilon)=\frac{\varepsilon}{D^2}\ln\frac{|\varepsilon^2-D^2|}{\varepsilon^2},\
\
\rho_0(\varepsilon)=\frac{|\varepsilon|}{D^2}\theta(D-|\varepsilon|)
\end{equation}
 with $D=v_Fk_c$ denoting a high-energy cutoff of the graphene bandwidth.
$k_c$ is chosen as to guarantee the conservation of the total
number of states in the Brillouin zone after the linearization of
the spectrum around the $K$ point, this is achieved  following the
Debye's prescription. We know from the selfenergy in Eq.
(\ref{faa2}) that introducing the impurities is behaving like
changing the Fermi level with the impurity concentration. A
similar effect, i.e. the Fermi level can be tuned from lying in
the conduction band to the gap, is experimentally demonstrated by
doping Ca in a topological insulator Bi$_{2}$Se$_{3}$
\cite{checkelsky}. The position of the impurity level is also
important since a resonance is clearly seen also as shown below.
In Eq.(\ref{jL5}), we further set the symmetrical voltage
division: $\mu_{L,R}=E_F\pm\frac{1}{2}eV$, and put $E_F=0$ in the
numerical calculations. The occupation of an impurity $\langle
n_\sigma\rangle$ which appears in the spin-dependent energy of
impurities should be treated in a self-consistent way via the
relation
\begin{equation}
\langle n_\sigma\rangle=-\frac{1}{\pi}\int_{-\infty}^{+\infty}
d\varepsilon f(\varepsilon)\text{Im}
G_f^r(\varepsilon),\label{selcon}
\end{equation}
 where
 \begin{equation}
\begin{array}{cll}
 G_{f}^{r}(\varepsilon)
&=&(1-\overline{g}_a^r\Sigma_d^r)[g_c^{-1}-\overline{g}_a^r(n_iV_0^2+g_c^{-1}\Sigma_d^r)]^{-1}
\end{array}\label{gff2}
\end{equation}
with $\Sigma_d^r(\varepsilon)=-\frac{i}{2}[\Gamma_{L}(\varepsilon)
+R \Gamma_{R}(\varepsilon)R^\dag]$. In terms of the result of
Eq.(\ref{jL5}), the TMR
can be obtained according to the
  conventional definition
\begin{equation}
TMR=\frac{I(0)-I(\pi)}{I(0)},\label{tmr}
\end{equation}
where $I(0,\pi)$ is the current flowing through the system in the
parallel (antiparallel) configuration.

\section{Bi- and tri-layer graphene}
Let us  consider a bilayer (trilyer) graphene by
adding a second (third) layer according to Bernal-type stacking
order (ABA). The inter-layer hopping energy is denoted by $t_p$. By a
 straightforward generalization of  the monolayer
 case we  obtain  the retarded Green's function
for the bilayer graphene as
\begin{equation}
\mathcal{G}_{2a}^{r}(\varepsilon) =\Lambda_2(\varepsilon)[1-
\Lambda_2(\varepsilon)\Sigma^r(\varepsilon)]^{-1}, \label{qbqb}
\end{equation}
where
\begin{equation}
\begin{array}{cll}
\Lambda_2(\varepsilon)
&=&\frac{1}{N}\sum\limits_\mathbf{q}\frac{-\varepsilon
v_F^2|\mathbf{q}|^2+\varepsilon^3- t_p^2\varepsilon}
{v_F^4|\mathbf{q}|^4-2\varepsilon^2 v_F^2|\mathbf{q}|^2+
\varepsilon^4- t_p^2\varepsilon^2}.
\end{array}\label{sigmaf1}
\end{equation}
For trilayer graphene we find
\begin{equation}
\mathcal{G}_{3a}^{r}(\varepsilon) =\Lambda_3(\varepsilon)[1-
\Lambda_3(\varepsilon)\Sigma^r(\varepsilon)]^{-1}, \label{qbqbb}
\end{equation}
where
\begin{equation}
\begin{array}{cll}
\Lambda_3(\varepsilon)
&=&-\frac{1}{N}\sum\limits_{\mathbf{q}}\frac{A_1v_F^4|\mathbf{q}|^4+B_1v_F^2|\mathbf{q}|^2+C_1}
{v_F^6|\mathbf{q}|^6+B_2v_F^4|\mathbf{q}|^4+C_2v_F^2|\mathbf{q}|^2+D_2}
\end{array}\label{sigmar}
\end{equation}
with
\[ A_1=\varepsilon, \: B_1=t_p^2\varepsilon -2\varepsilon^3,\]
\[C_1=\varepsilon^5 -2t_p^2\varepsilon^3, \,B_2=-3\varepsilon^2,\,
C_2=-2t_p^2\varepsilon^2 +3\varepsilon^4,\]  and \[D_2=-\varepsilon^6
+2t_p^2\varepsilon^4.\]
 Substituting Eqs.(\ref{qbqb}) and (\ref{qbqbb}) in
Eq.(\ref{jL5}), we obtain the electric current through the bi- and
tri-layer graphene with impurity atoms, and also the TMR according
to Eq.(\ref{tmr}). The summation over $\mathbf{q}$ in
Eqs.(\ref{sigmaf1}) and (\ref{sigmar}) 
 may be  performed by taking the continuum
limit with a  cutoff $D$ and expanding the
integrand in terms of  partial fractions.
 The final results have the explicit form
\begin{equation}
\begin{array}{cll}
\Lambda_2(\varepsilon) &=&-\frac{1}{D^2}\{
\frac{\text{sgn}(\varepsilon)t_p}{2}
\ln|\frac{(D^2-x_1)x_2}{(D^2-x_2)x_1}|\\
&&+\frac{\varepsilon}{2}\ln |\frac{(D^2-x_1)(D^2-x_2)}{x_1x_2}|\}
-i \frac{\pi }{D^2} \{\frac{\text{sgn}(\varepsilon)t_p}{2}\\
&&\times[\text{sgn}(\frac{d x_1}{d\varepsilon})\theta(0<x_1<D^2)\\
&&-\text{sgn}(\frac{d x_2}{d\varepsilon})\theta(0<x_2<D^2)]\\
&&+\frac{\varepsilon}{2}[\text{sgn}(\frac{d
x_1}{d\varepsilon})\theta(0<x_1<D^2)\\
&&+\text{sgn}(\frac{d x_2}{d\varepsilon})\theta(0<x_2<D^2)]\}
,\label{sigmaf}
\end{array}
\end{equation}
where $\theta(x)$ is the step function, and $
x_{1,2}=\varepsilon^2\pm t_p|\varepsilon| $. For  $\Delta=(2B_2^3
-9B_2C_2+27D_2)^2+4(-B_2^2+3C_2)^3\geq 0$,
\begin{equation}
\begin{array}{cll}
\Lambda_3(\varepsilon) &=&-\frac{1}{D^2}
\{[A_1+\frac{A_1(x_2+x_3)x_1+B_1x_1
-A_1x_2x_3+C_1}{(x_2-x_1)(x_3-x_1)}]\\
&&\times\ln|\frac{D^2-x_1}{x_1}|
+\frac{A_1(x_2+x_3)+B_1}{\sqrt{x_2x_3-(x_2+x_3)^2/4}}\\
&&\times(\arctan\frac{D^2-(x_2+x_3)/2}{\sqrt{x_2x_3-(x_2+x_3)^2/4}}\\
&&+\arctan\frac{(x_2+x_3)/2}{\sqrt{x_2x_3-(x_2+x_3)^2/4}})\\
&&+\frac{A_1(x_2+x_3)x_1+B_1x_1
-A_1x_2x_3+C_1}{(x_2-x_1)(x_3-x_1)}\\
&&\times[\frac{-1}{2}\ln\frac{D^2-(x_2+x_3)+x_2x_3}{x_2x_3}\\
&&+\frac{(x_2+x_3)/2-x_1}{2\sqrt{x_2x_3-(x_2+x_3)^2/4}}\\
&&\times(\arctan\frac{D^2-(x_2+x_3)/2}{\sqrt{x_2x_3-(x_2+x_3)^2/4}}\\
&&+\arctan\frac{(x_2+x_3)/2}{\sqrt{x_2x_3-(x_2+x_3)^2/4}})]\}\\

&&-i\text{sgn}(\frac{dx_1}{d\omega})\theta(0<x_1<D^2)\frac{\pi
V_f^2}{D^2} \\
&&\times[A_1+\frac{A_1(x_2+x_3)x_1+B_1x_1
-A_1x_2x_3+C_1}{(x_2-x_1)(x_3-x_1)}],
\end{array}\label{tsigma1}
\end{equation}
where
\begin{equation}
\begin{array}{cll}
x_1 &=&-\frac{B_2}{3}+\frac{1}{2^{1/3}}\frac{1}{3}\{-2B_2^3
+9B_2C_2-27D_2\\
&&+\sqrt{(2B_2^3
-9B_2C_2+27D_2)^2+4(-B_2^2+3C_2)^3}\}^{\frac{1}{3}}\\
&&+\frac{1}{2^{1/3}3}\{-2B_2^3 +9B_2C_2-27D_2\\
&&-\sqrt{(2B_2^3
-9B_2C_2+27D_2)^2+4(-B_2^2+3C_2)^3}\}^{\frac{1}{3}},
\end{array}
\end{equation}
\begin{equation}
\begin{array}{cll}
x_{2,3}&=&-\frac{B_2}{3}+\frac{-\frac{1}{2}-i\frac{\sqrt{3}}{2}}{3}\frac{1}{2^{1/3}}\{-2B_2^3
+9B_2C_2-27D_2\\
&&\pm\sqrt{(2B_2^3
-9B_2C_2+27D_2)^2+4(-B_2^2+3C_2)^3}\}^{\frac{1}{3}}\\
&&+\frac{-\frac{1}{2}+i\frac{\sqrt{3}}{2}}{3}\frac{1}{2^{1/3}}\{-2B_2^3
+9B_2C_2-27D_2\\
&&\mp\sqrt{(2B_2^3
-9B_2C_2+27D_2)^2+4(-B_2^2+3C_2)^3}\}^{\frac{1}{3}}.
\end{array}
\end{equation}
For $\Delta<0$,
\begin{equation}
\begin{array}{cll}
\Lambda_3(\varepsilon) &=&-\frac{1}{D^2}\{
[A_1+\frac{[A_1(x_2+x_3)+B_1]x_1-A_1x_2x_3+C_1}
{(x_2-x_1)(x_3-x_1)}]\\
&&\times\ln|\frac{D^2-x_1}{x_1}| -
[\frac{A_1(x_2+x_3)+B_1}{x_3-x_2}\\
&&+\frac{[A_1(x_2+x_3)+B_1]x_1-A_1x_2x_3+C_1}
{(x_2-x_1)(x_3-x_2)}]\ln|\frac{D^2-x_2}{x_2}|\\
&&+
[\frac{A_1(x_2+x_3)+B_1}{x_3-x_2}\\
&&+\frac{[A_1(x_2+x_3)+B_1]x_1-A_1x_2x_3+C_1}
{(x_2-x_1)(x_3-x_2)}\\
&&-\frac{[A_1(x_2+x_3)+B_1]x_1-A_1x_2x_3+C_1}
{(x_2-x_1)(x_3-x_1)}]\\
&&\times\ln|\frac{D^2-x_3}{x_3}|\} -i\frac{\pi
V_f^2}{D^2}\{\text{sgn}(\frac{dx_1}{d\omega})\theta(0<x_1<D^2)\\
&&\times[A_1+\frac{[A_1(x_2+x_3)+B_1]x_1-A_1x_2x_3+C_1}
{(x_2-x_1)(x_3-x_1)}]\\
&&-\text{sgn}(\frac{dx_2}{d\omega})\theta(0<x_2<D^2)
[\frac{A_1(x_2+x_3)+B_1}{x_3-x_2}\\
&&+\frac{[A_1(x_2+x_3)+B_1]x_1-A_1x_2x_3+C_1}
{(x_2-x_1)(x_3-x_2)}]\\
&&+\text{sgn}(\frac{dx_3}{d\omega})\theta(0<x_3<D^2)
[\frac{A_1(x_2+x_3)+B_1}{x_3-x_2}\\
&&+\frac{[A_1(x_2+x_3)+B_1]x_1-A_1x_2x_3+C_1}
{(x_2-x_1)(x_3-x_2)}\\
&&-\frac{[A_1(x_2+x_3)+B_1]x_1-A_1x_2x_3+C_1}
{(x_2-x_1)(x_3-x_1)}]\},
\end{array}\label{tsigma2}
\end{equation}
where
\begin{equation}
\begin{array}{cll}
x_1 &=&-\frac{B_2}{3}+\frac{2\sqrt{B_2^2-3C_2}}{3}\cos(
\frac{\arccos T}{3}),\\
x_2 &=&-\frac{B_2}{3}+\frac{2\sqrt{B_2^2-3C_2}}{3}\cos(
\frac{2\pi+\arccos T}{3}),\\
x_3 &=&-\frac{B_2}{3}+\frac{2\sqrt{B_2^2-3C_2}}{3}\cos(
\frac{4\pi+\arccos T}{3}),\\
 T&=&-\frac{2(B_2^2-3C_2)B_2
-3(B_2C_2-9D_2)}{2(B_2^2-3C_2)^{\frac{3}{2}}}.
\end{array}
\end{equation}
 To determine $\langle n_\sigma\rangle$, we
need to calculate the impurity Green's function
$G_{i=2,3;f}^{\sigma\sigma',r}(t-t')
=-i\theta(t-t')\langle\{f_\sigma(t),f_{\sigma'}^\dag(t')\}\rangle$
for the bilayer and trilayer graphene. Using the equation of
motion  method we find
 \begin{equation}
\begin{array}{cll}
 G_{i,f}^{r}(\varepsilon)
&=&(1-\Lambda_i(\varepsilon)\Sigma_d^r)[g_c^{-1}-\Lambda_i(\varepsilon)(n_iV_0^2+g_c^{-1}\Sigma_d^r)]^{-1}
\end{array}.\label{ff2gfqa}
\end{equation}
Substituting Eq.(\ref{ff2gfqa}) in Eq.(\ref{selcon}), we  find $\langle n_\sigma\rangle$
 self-consistently.

\begin{figure}[h]
\center
\includegraphics[width=1.12\columnwidth ]{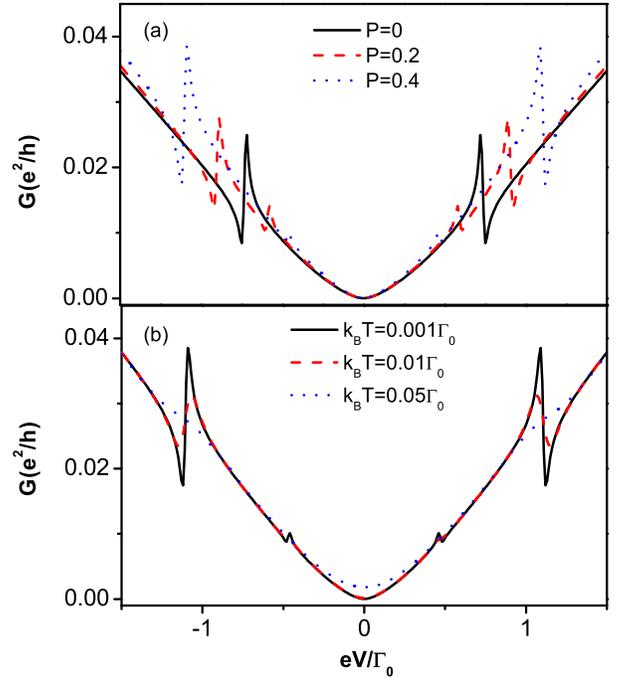}
\caption{ (color online) The bias dependence of the differential
conductance $G$ for different polarizations $P$ at
$T=0.001\Gamma_0$ (a) and for different  temperatures $T$ at
$P=0.4$ (b) for a single layer graphene junction and for parallel
configuration of the electrodes magnetizations. The other
parameters are taken as $V_f=3\Gamma_0,n_i=0.1,
\varepsilon_0=0.1\Gamma_0,U=\Gamma_0,D=5\Gamma_0$, where
$\Gamma_0$ stands for the coupling between the scattering region
and
 the electrodes.}\label{fig2}
\end{figure}

\section{Numerical analysis}
To illustrate the nature of the spin-dependent transport
we performed numerical calculations  assuming the linewidth function
$\Gamma_\alpha^\sigma(\varepsilon)$ to be independent of the energy
within the wide band approximation. Introducing the degree of the spin
polarizations of the left and the right electrodes,$P_L$
and $P_R$, and assuming that the two electrodes are made of the
same material, i.e. $P_L=P_R=P$, we can write
$\Gamma_L^{\uparrow\downarrow}=\Gamma_R^{\uparrow\downarrow}=\Gamma_0(1\pm
P)$ where $\Gamma_0$ describes the coupling between the graphene
and the electrodes in absence of an internal magnetization, and is taken
as the energy scale in the following numerical calculation.

Fig.\ref{fig2}(a) shows the bias dependence of the differential
conductance $G=dI/dV$ in a single layer graphene system for the
different $P$ under the parallel configuration of the electrodes
magnetizations. When $P=0$ (nonmagnetic electrodes), the
differential conductance possesses a single resonant peak as the
bias voltage increases (or decreases) from zero up to few
$\Gamma_0$ (or $-\Gamma_0$). This peak
corresponds to a resonant tunneling through the renormalized spin-dependent impurity level 
$\varepsilon_0+Un_\sigma$. Thereby, this situation reflects the
fact that the impurity is not magnetized in the absence of the
internal magnetization of the electrodes, i.e.
$n_\uparrow=n_\downarrow$. When  $P\neq 0$, the aforementioned
single peak in the conductance splits into two peaks indicating
the positions of the resonant tunneling at
$\varepsilon_0+Un_\uparrow$ and $\varepsilon_0+Un_\downarrow$ and
signaling the form of the localized magnetic moment of the
impurity. This behavior is due to the fact
 that the exchange splitting
of the density of state (DOS) in the ferromagnetic leads acts as
an effective magnetic field with values well larger than those of
an externally applied field\cite{Pasupathy2004sci,zhu2008pla}.
This effective field acts on the graphene including the impurity
atoms and influences strongly the spin states of the impurities.
With increasing $P$ the
 peaks split further way from each other signifying that
the polarization in the ferromagnetic leads enhances the localized
magnetic moment.
 With increasing $P$, the peak at the higher
voltage is enhanced and that at the lower voltage is suppressed.
This non-monotonic dependence of the peak amplitude on the
polarization of the electrodes is completely different from the
ferromagnet-quantum dot-ferromagnet(FM-QD-FM) system
\cite{sergueevprb2002,muprb2006,zzgprb2004}; its origin stems from
the linear DOS of graphene.
\begin{figure}[h]
\center
\includegraphics[width=1.15\columnwidth ]{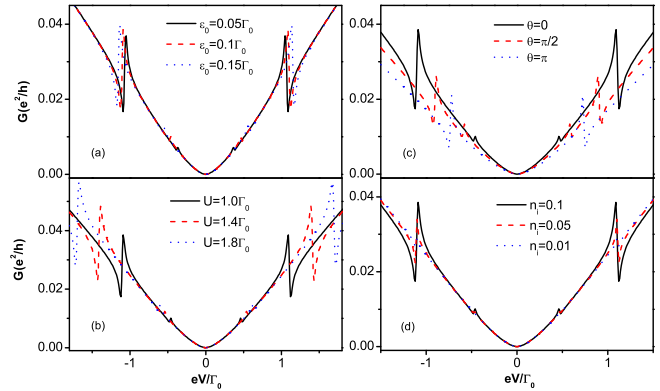}
\caption{(color online)The bias dependence of the differential
conductance $G$ for different energies $\varepsilon_0$ of the
impurity  level  at $U=\Gamma_0$, $\theta=0$ and $n_i=0.1$ (a),
for different interaction strengths $U$ at
$\varepsilon_0=0.1\Gamma_0$, $\theta=0$ and $n_i=0.1$ (b), and for
different angles $\theta$ at $\varepsilon_0=0.1\Gamma_0$,
$U=\Gamma_0$ and $n_i=0.1$ (c) as well as for the different
impurity concentrations $n_i$ at $\varepsilon_0=0.1\Gamma_0$,
$U=\Gamma_0$ and $\theta=0$ (d). A
 single layer of graphene is considered. The other parameters are taken the
same as those of Fig.\ref{fig2}.}\label{fig3}
\end{figure}
The dependence of the differential conductance on the bias for
different temperatures in a parallel configuration is shown in
Fig. \ref{fig2}(b) for a single layer graphene device. With
increasing the temperature, the resonant peaks in the differential
conductance decrease and almost vanish at larger $T$. This
temperature dependence of the conductance peaks is similar to that
of a noninteracting, single-particle resonance in the
multi-channel model. The mechanism for this destruction of the
peak is that at high temperatures not all electron states of the
low-lying subbands are fully occupied  due to the occupation  of
the next
 subbands\cite{vanprb1991}. However, near $V=0$, the conductance
increases when the temperature is raised. This characteristic
feature is the same as that in the clean undoped graphene systems 
\cite{ding2009prb}. The reason is the graphene acts as a
barrier at the zero energy point. Near zero-bias voltage, the
increase of the thermally excited electrons with the temperature
enhances the conductance.

\begin{figure}[h]
\includegraphics[width=1.12\columnwidth ]{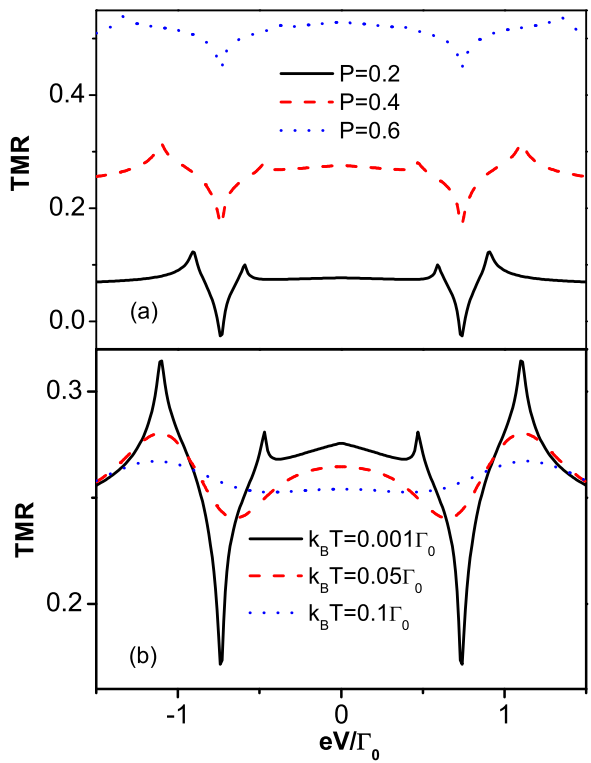}
\caption{(color online) The bias dependence of the TMR for different degrees of polarization
$P$ at $T=0.001\Gamma_0$ (a) and for different temperatures $T$ at
$P=0.4$ (b) for a single layer graphene tunnel junction. The other
parameters are taken the same as those of
Fig.\ref{fig2}.}\label{fig4}
\end{figure}
The bias dependence of the differential conductance of a single
layer graphene device for
different  $\varepsilon_0$ and 
different interaction $U$ is
shown in Fig. \ref{fig3}(a) and (b). With increasing
$\varepsilon_0$ and $U$, the resonant peaks in the conductance shift
towards  large bias voltages, and the peaks amplitudes
become large which is also related to the linear DOS of the
graphene. In particular, one can find that the interval between
the conductance peaks increases with the interaction $U$, in
contrast it decreases with $\varepsilon_0$. 
This is because  the DOS  around the impurity energy level is
enhanced as $\varepsilon_0$ grows, which decreases the localized
magnetic moment on the impurity
\cite{Uchoaprl2008,{dingjpcm2009}}. Fig.\ref{fig3}(c) shows for a
single layer graphene tunnel junction the bias dependence of the
differential conductance when varying the mutual angle $\theta$
between the electrodes magnetization direction. With increasing
$\theta$, the split structure of the resonant peaks in the
differential conductance is washed out, and
 eventually the two peaks merge into a single peak at $\theta=\pi$,
meaning that the impurity loses \textit{its} magnetism in this
situation. This behavior suggests that the magnetization
electrodes can be used as a valve device to open or close the
magnetism of impurity atoms in graphene. Similar phenomena has
also been observed in the FM-QD-FM system
\cite{Pasupathy2004sci,zhangarxiv2002}, where the magnetic moment
in the quantum dot is induced by different spin-dependent
tunneling rates between the two magnetic electrodes. In our case,
the impurity atom immerses in the Dirac fermions sea.  Its
magnetic moment is turned on via a mediation of the itinerant
massless Dirac fermion not via a direct tunneling
\cite{Uchoaprl2008,dingjpcm2009}. This shows new features as
discussed above and below. In Fig. \ref{fig3}(d), we plot the
differential conductance as an function of bias voltage for
different impurity concentrations in a single layer graphene
device. One can easily observe that the resonant peaks diminish
with decreasing the impurity concentration, however their
positions do not change. This suggests that the magnetization is
maintained although with descendent effect of the impurities on
transport in graphene at lower concentration since the information
of the magnetizations of the electrodes is transmitted by the
Dirac fermions in the same way.
\begin{figure}[h]
\includegraphics[width=1.12\columnwidth ]{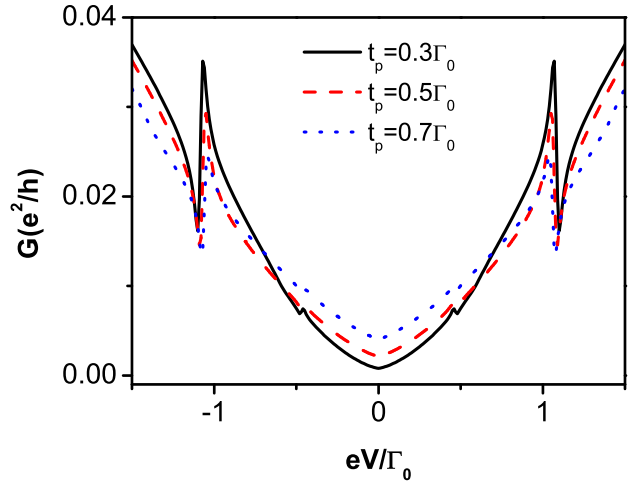}
\caption{(color online) The bias dependence of the differential conductance $G$
for different interlayer couplings $t_p$ at $T=0.001\Gamma_0$ and
$P=0.4$ for the bilayer graphene system. The other parameters are
taken the same as those of Fig.\ref{fig2}.}\label{fig5}
\end{figure}

The bias dependence of the TMR, defined in Eq.(\ref{tmr}), for
different polarizations $P$ and the different temperatures $T$ in
a single layer graphene junction is shown in Fig.\ref{fig4}. A
small hump at zero bias corresponds to an enhanced
magnetoresistance due to the result of a nontrivial combined
effect of graphene and conventional spin-valve properties
\cite{ding2009prb}. The existence of the impurity suppresses the
amplitude of this structure since
 the impurity influence on the DOS extends to the Dirac
point\cite{nilssprl2007}. Additionally, it is found that the TMR
as a function of the bias voltage exhibits two pronounced dips at
lower and  higher bias. The latter one is particularly interesting
as the TMR there may change  sign  from positive to negative for
small values of the polarization parameters. To account for this
behavior one should take into account the angular dependence of
the
 conductance from Fig.\ref{fig3}(c) which shows clearly
the change of the positions and the heights of the peak with the
angle $\theta$, leading to the appearance of the negative TMR. For
large bias (see Fig. \ref{fig4}(b)), the graphene system with the
impurity atoms behaves like that in the absence of the
impurities\cite{ding2009prb}, where TMR hardly changes with
temperature $T$. This phenomena suggests that the presence of the
impurity atoms does not affect the magnetoresistance at  high bias
voltages. However, for a low bias voltage, TMR displays  many
important features: With increasing temperatures the hump at zero
bias diminishes since the increase in the electrical current from
the contribution of thermally excited electrons is faster for the
antiparallel configuration. Additionally, the dip in TMR decreases
with $T$. Therefore, due to their combinations, TMR near the zero
bias voltage displays a broad peak instead of the "W"-shape
feature with increasing the temperature $T$, and eventually
develops into a broad dip.

\begin{figure}[h]
\includegraphics[width=1.12\columnwidth ]{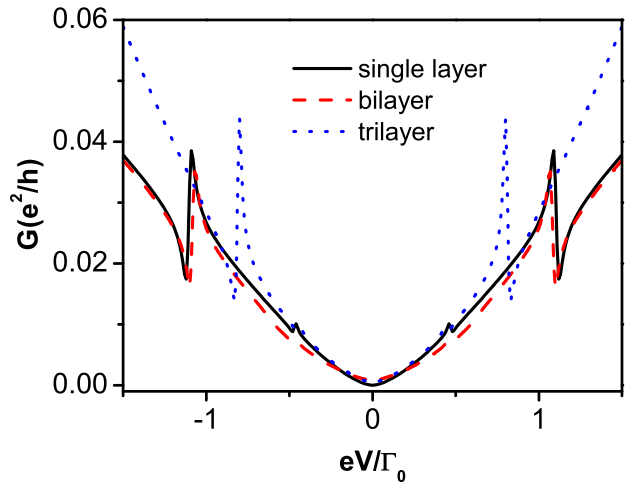}
\caption{(color online) The bias dependence of the differential conductance $G$
for different layer graphene at $T=0.001\Gamma_0$, $P=0.4$ and
$t_p=0.3\Gamma_0$. The other parameters are taken the same as
those of Fig.\ref{fig2}.}\label{fig6}
\end{figure}
Fig.\ref{fig5} shows the bias dependence of the differential
conductance in a bilayer graphene
system. When the interlayer coupling grows, the conductance at
zero bias increases, while the resonant peak diminishes. The
explanation for this phenomenon is as follows:  the quasiparticles
in the bilayer graphene have a peculiar nature with features akin
both to Dirac and to conventional fermions. The contribution of
conventional fermions originates from the interlayer coupling that
supports a metallic bilayer graphene. Thus the increase of the
interlayer coupling enhances the zero bias conductance in the
bilayer graphene. Furthermore, 
the metallic behavior
due to the interlayer coupling enhances the screening of the
impurity \cite{diviprb1984,peres}, which leads to the decrease of
the transmission probability of the electrons, thus suppressing
the resonant tunneling through the energy level
$\varepsilon_0+n_\sigma U$. The bias dependence of the
differential conductance for the different layer graphene  devices %
is shown in Fig. \ref{fig6}. The differential conductance for all
the graphene layers displays two peaks corresponding to the
resonant tunneling through the energy level
$\varepsilon_0+n_\sigma U$. It is suggested that the impurity
magnetic moment also exists for the bilayer and trilayer graphene
when the ferromagnetic electrodes are introduced. However,  the
distance of the two peaks decreases with increasing the number of
graphene layers. This result stems from the fact that the impurity
magnetic moment decreases with increasing the layer number
\cite{dingjpcm2009}.

\section{Summary}
In conclusion, using Keldysh's nonequilibrium Green's function
method, we study theoretically the spin dependent transport
through  a tunnel junction made of few layers of  graphene with
impurity atoms. It is found that when the electrodes magnetization
vectors are parallel,
 the bias dependence of the differential conductance
exhibits two resonant peaks as the bias voltage increases.
We assign this behavior to  the formation of a magnetic moment
on the impurity  due to the electrodes spin polarization.
With increasing the mutual angle $\theta$ between the vectors of the
 two electrodes magnetization, the
distance of the two resonant peaks in the differential conductance
diminishes, and they merge eventually  into a single peak at  $\theta=\pi$.
 The reason is that  the  induced
magnetic moment on the  impurity decreases with $\theta$
and  vanishes for $\theta=\pi$. This
 features may be exploited to develop a
magnetic valve device that operates based on the presence  or the absence of the magnetism on the
impurity atom. Furthermore, due to the  $\theta$-dependence
 of the impurity magnetic
moment, the TMR may change  sign
for small values of the polarization parameters.

\begin{acknowledgments}{
The work of K.H.D. and Z.H.Z. was supported by the National
Natural Science Foundation of China (Grant Nos. 10904007 and
60771059), the Natural Science Foundation of Hunan Province, China
(Grant No. 08JJ4002 ), and the construct program of the key
discipline in Changsha University of Science and Technology,
China. J.B. and Z.G.Z. were supported by the cluster of excellence
"Nanostructured Materials" of the state Saxony-Anhalt, and the DFG, Germany.}
\end{acknowledgments}

\end{document}